\documentclass[aps,floats,showpacs,nofootinbib]{revtex4}

\usepackage{pifont,color,multirow,graphicx,mathrsfs,makeidx,epsfig,fancyhdr,fancybox,calc,
amsmath,amsfonts,amssymb,amsthm,latexsym,comment,appendix}

\usepackage[latin1]{inputenc}

\newcommand{\nc}{\newcommand}
\nc{\bb}{\begin{equation}} \nc{\ee}{\end{equation}}
\nc{\ug}{\; = \;} \nc{\tr}{\triangle} \nc{\rd}{{\rm d}}
\nc{\R}{{\rm {I\!R}}} \nc{\vs}{\vspace*} {\nc{\un}{1\!\!1}
\nc{\0}{\vs*{0.5cm}} \nc{\h}{\hspace*{0.5cm}}
\nc{\vare}{\varepsilon} \nc{\dis}{\displaystyle}
\nc{\eee}{\end{document}}
\nc{\modl}{\|} \nc{\modr}{\|}    
\nc{\um}{\frac{1}{2}} \nc{\rdt}{\rd t}
\nc{\pa}{\partial} \nc{\erm}{{\rm e}}
\nc{\Mbar}{\overline{M}} \nc{\lbar}{\overline{l}}
\nc{\Lc}{{\cal L}} \nc{\Sc}{{\cal S}} \nc{\const}{{\rm constant}}
\nc{\impl}{{ \ \ \ \Rightarrow \ \ \ }}
\nc{\Impl}{{ \ \ \ \Longrightarrow \ \ \ }}
\nc{\oi}{{0i}} \nc{\io}{{i0}} \nc{\ok}{{0k}} \nc{\ko}{{k0}}
\nc{\psib}{\overline{\psi}} \nc{\psid}{{\psi^{\dagger}}}
\nc{\munu}{{\mu\nu}} \nc{\C}{{\rm I\!\!\!C}}
\nc{\vecna}{\mbox{\boldmath $\nabla$}}
\nc{\Hc}{{\cal H}} \nc{\Lcn}{{\cal L}^{(n)}}
\nc{\Lczero}{{\cal L}^{(0)}} \nc{\Lcuno}{{\cal L}^{(1)}}  
\nc{\Lcdue}{{\cal L}^{(2)}} \nc{\Lctre}{{\cal L}^{(3)}}  
\nc{\ga}{\gamma} \nc{\al}{\alpha} \nc{\gao}{\gamma^0} \nc{\pmu}{p^\mu} 
\nc{\vmu}{v^\mu} \nc{\icsmu}{x^\mu} \nc{\gmu}{\ga^\mu} \nc{\gnu}{\ga^\nu}

\nc{\Ebf}{\mbox{\boldmath $E$}} \nc{\Hbf}{\mbox{\boldmath $H$}}
\nc{\Vbf}{\mbox{\boldmath $V$}} \nc{\Fbf}{\mbox{\boldmath $F$}}
\nc{\Wbf}{\mbox{\boldmath $W$}} \nc{\lbf}{\mbox{\boldmath $l$}}
\nc{\xbf}{\mbox{\boldmath $x$}} \nc{\ubf}{\mbox{\boldmath $u$}}
\nc{\vbf}{\mbox{\boldmath $v$}} \nc{\wbf}{\mbox{\boldmath $w$}}
\nc{\jbf}{\mbox{\boldmath $j$}} \nc{\mubf}{\mbox{\boldmath $\mu$}}
\nc{\sigbf}{\mbox{\boldmath $\sigma$}} \nc{\pibf}{\mbox{\boldmath $\pi$}}
\nc{\gabf}{\mbox{\boldmath $\gamma$}} \nc{\albf}{\mbox{\boldmath $\alpha$}}
\nc{\Xbf}{\mbox{\boldmath $X$}} \nc{\kbf}{\mbox{\boldmath $k$}}
\nc{\qbf}{\mbox{\boldmath $q$}} \nc{\Qbf}{\mbox{\boldmath $Q$}}
\nc{\abf}{\mbox{\boldmath $a$}} \nc{\bbf}{\mbox{\boldmath $b$}}
\nc{\sbf}{\mbox{\boldmath $s$}} \nc{\rbf}{\mbox{\boldmath $r$}}
\nc{\Lbf}{\mbox{\boldmath $L$}} \nc{\imp}{\mbox{\boldmath $p$}}

\nc{\dddov}{{\stackrel{\ldots}{v}}} \nc{\dox}{\dot{x}} \nc{\ddox}{\ddot{x}}
\nc{\dddox}{{\stackrel{\ldots}{x}}} \nc{\dopi}{\dot{\pi}} \nc{\dop}{\dot{p}}
\nc{\dov}{\dot{v}} \nc{\ddov}{\ddot{v}} \nc{\doa}{\dot{a}} \nc{\ddoa}{\ddot{a}}
\nc{\dddoa}{{\stackrel{\ldots}{a}}} \nc{\ddddoa}{{\stackrel{....}{a}}}
\nc{\ddddov}{{\stackrel{....}{v}}} \nc{\doX}{\dot{X}} \nc{\ddoX}{\ddot{X}}
\nc{\doS}{\dot{S}} \nc{\ddoS}{\ddot{S}} \nc{\doJ}{\dot{J}}
\nc{\doL}{\dot{L}} \nc{\ddoL}{\ddot{L}} \nc{\doq}{\dot{q}} \nc{\ddoq}{\ddot{q}}
\nc{\doxbf}{\dot{\xbf}} \nc{\doXbf}{\dot{\Xbf}} \nc{\ddoXbf}{\ddot{\Xbf}}
\nc{\ddoxbf}{\ddot{\xbf}} \nc{\doabf}{\dot{\abf}} \nc{\doqbf}{\dot{\qbf}}
\nc{\dopibf}{\dot{\pibf}} \nc{\ddoabf}{\ddot{\abf}}
\nc{\dddoabf}{\mbox{\boldmath $\dddoa$}} \nc{\ddddoabf}{\mbox{\boldmath $\ddddoa$}}

\nc{\dtau}{{\rd\tau}} \nc{\dt}{{\rd t}} \nc{\pp}{{p_\mu v^\mu}} 
\nc{\ppo}{{\po_\mu \gamma^\mu}} \nc{\SCM}{_{\star}} \nc{\vi}{{v^{(\rm i)}}} 
\nc{\aii}{{a^{(\rm 2i)}}} \nc{\M}{{\rm I\!M\!I}} \nc{\gaoinv}{{{\gao}^{-1}}} \nc{\voinv}{{v_0}^{-1}} \nc{\voinvo}{\widehat{\voinv}}
\nc{\vbfo}{\widehat{\vbf}} \nc{\abfo}{\widehat{\abf}}
\nc{\abfod}{\widehat{\abf}^\dagger} \nc{\aoh}{\widehat{a}} \nc{\vo}{\widehat{v}}
\nc{\qo}{\widehat{q}} \nc{\pio}{\widehat{\pi}}
\nc{\So}{\widehat{S}} \nc{\Lo}{\widehat{L}} \nc{\Go}{\widehat{G}}
\nc{\sbfo}{\widehat{\sbf}} \nc{\aod}{{\widehat{a}}^\dagger}
\nc{\impo}{\widehat{\imp}} \nc{\po}{\widehat{p}}
\nc{\Xio}{\widehat{\Xi}} \nc{\Jo}{\widehat{J}}
\nc{\doimpo}{\mbox{\boldmath ${{\stackrel{.}{\impo}}}$}}
\nc{\doabfo}{\widehat{\doabf}} \nc{\qbfo}{\widehat{\qbf}}
\nc{\pibfo}{\widehat{\pibf}}  \nc{\Hco}{\widehat{\Hc}}
\nc{\doqbfo}{\widehat{\doqbf}} \nc{\dopibfo}{\widehat{\dopibf}}
\nc{\ddoqbf}{\ddot{\qbf}}  \nc{\doao}{\dot{\aoh}}
\nc{\ddoqbfo}{\widehat{\ddoqbf}} \nc{\doqo}{\widehat{\doq}} 
\nc{\dopio}{\widehat{\dopi}} \nc{\ddoqo}{\widehat{\ddoq}} 
\nc{\Xbfo}{\widehat{\Xbf}} \nc{\doxo}{\widehat{\dox}} 
\nc{\doXo}{\widehat{\doX}} \nc{\xo}{\widehat{x}} \nc{\Xo}{\widehat{X}}
\nc{\ddoXo}{\widehat{\ddoX}} \nc{\xbfo}{\widehat{\xbf}}
\nc{\doxbfo}{\widehat{\doxbf}} \nc{\doXbfo}{\widehat{\doXbf}}
\nc{\ddoXbfo}{\widehat{\ddoXbf}} \nc{\dopo}{\widehat{\dop}}
\nc{\dovo}{\widehat{\dov}} \nc{\ddovo}{\widehat{\ddov}}

\nc{\Sigbf}{\mbox{\boldmath $\Sigma$}}
\nc{\Orm}{{\rm O}\!\!\!\!\!{\rm O}} \nc{\qt}{\tilde{q}}
\nc{\ddoR}{\ddot{R}} \nc{\ddddoR}{{\stackrel{....}{R}}}
\nc{\nnmuno}{{NNM$^{(1)}$}\,} \nc{\omegao}{\omega_0}
\nc{\rmi}{{\rm i}} \nc{\rmg}{_{{\footnotesize{\rm G}}}}
\nc{\Wt}{{\widetilde{W}}} \nc{\doWt}{\dot{\Wt}}
\nc{\Wto}{\widehat{\Wt}} \nc{\xm}{\overline{x}}
\nc{\xmo}{\widehat{\overline{x}}}
\nc{\xbfmo}{\widehat{\overline{\xbf}}}
\nc{\CoMF}{_{\rm {\footnotesize CMF}}} \nc{\cmf}{_{\star}}
\nc{\lqq}{\left[} \nc{\rqq}{\right]}
\nc{\lt}{\left(} \nc{\rt}{\right)}
\nc{\ik}{{ik}} \nc{\doSo}{\widehat{\doS}}

\nc{\nn}{\noindent}
\nc{\interitem}{\vspace*{-0.5 cm}\setlength{\itemsep}{1pt}\setlength{\parskip}{0pt}}
\nc{\intraitem}{\vspace*{-0.2 cm}\setlength{\itemsep}{1pt}\setlength{\parskip}{0pt}}
\nc{\traitem}{\setlength{\itemsep}{1pt}\setlength{\parskip}{-2pt}}
\nc{\fineitem}{\vspace*{-0.5cm}}
\nc{\e}{\`e } \nc{\oo}{\`o } \nc{\uu}{\`u } \nc{\ac}{\`a } \nc{\ii}{\`{\i \ }}
\nc{\Pibf}{\mbox{\boldmath $\Pi$}} \nc{\Hcn}{{\cal H}^{(n)}} \nc{\omz}{\omega_0}
\nc{\kb}{\overline{k}} \nc{\alpt}{{\widetilde{\alpha}}} \nc{\alm}{{\alpha_m^\mu}}
\nc{\alter}{{\alpt_m^\mu}} \nc{\almd}{{\alpha^{\dagger\mu}_m}}
\nc{\altd}{{\alpt^{\dagger\mu}_m}} \nc{\almbf}{{\albf_m}}
\nc{\altbf}{{{\widetilde{\albf}}_m}} \nc{\almdbf}{{\albf^{\dagger}_m}}
\nc{\altdbf}{{\widetilde{\albf}^{\dagger}_m}} 
\nc{\dosbf}{\mbox{\boldmath ${{\stackrel{.}{s}}}$}} \nc{\Ec}{E} \nc{\lpt}{\left(}
\nc{\alphao}{\widehat{\alpha}} \nc{\betao}{\widehat{\beta}}
\nc{\dbf}{\mbox{\boldmath $d$}} \nc{\okbf}{\overline{\mbox{\boldmath $k$}}}
\nc{\Pio}{\widehat{\Pi}} \nc{\nnmdue}{{NNM$^{(2)}$}\,} \nc{\lht}{\hat{l}}
\nc{\np}{n_p} \nc{\psis}{{\psi^{\star}}} \nc{\psibf}{\mbox{\boldmath $\psi$}}  
\nc{\up}{u_p} \nc{\norm}{{\frac{1}{\sqrt{2\vare}}}} \nc{\para}{^{\parallel}} 
\nc{\orto}{^{\perp}} \nc{\tz}{\tau_0}
\nc{\psidbf}{\mbox{\boldmath $\psid$}} \nc{\chis}{\chi^{\star}} \nc{\Psib}{\overline{\Psi}}
\nc{\aoo}{\widehat{a}} \nc{\bo}{\widehat{b}}
\nc{\aop}{\aoo_{{\footnotesize \imp}}} \nc{\bop}{\bo_{{\footnotesize \imp}}}
\nc{\aopd}{\aoo_{{\footnotesize \imp}}^\dagger}
\nc{\bopd}{\bo_{{\footnotesize \imp}}^\dagger}
\nc{\intp}{\int\rd^3p} \nc{\LcN}{{\cal L}^{(N)}} 

\begin{document}

\title{Helicity-0 spinning particles\footnote{Work partially supported by INFN and MURST}}

\author{G. Salesi$^{\rm a,b}$\footnote{Corresponding author; e-mail: salesi@unibg.it}}

\author{L. Deleidi$^{\rm a}$}

\affiliation{\ \\ \mbox{$^{\rm a}$Dipartimento di Ingegneria e Scienze Applicate, Universit\`a di Bergamo, viale Marconi 5, Dalmine, Italy}}

\affiliation{\mbox{$^{\rm b}$Istituto Nazionale di Fisica Nucleare, Sezione di Milano, via Celoria 16, Milan, Italy}}

\begin{abstract} 

\vspace*{0.1cm}
\nn We show that a self-consistent classical theory of the spin, based on a 
very general Lagrangian extending the Newtonian dynamics, does predict the special case 
of helicity-0 particles, which at the same time are endowed with nonzero spin and zero 
intrinsic angular momentum. 

\noindent

\end{abstract}

\maketitle

\section{Introduction}

\nn Besides the ordinary cases of spinless scalars and spinnning spinors and vectors, it exists another exotic class of elementary particles: the helicity-0 particles (HZPs), which are spinning, but with zero
helicity and vanishing conserved component of the spin vector. 
HZPs have occasionally been associated to longitudinal excitations of vector fields found in string theory and in some dark-matter and dark-energy 
cosmological theories (sometimes recognized as ``ghost'' fields or ``aether''-like fields).
In particular, both longitudinal and trasversal components of vector fields with a nonzero covariant divergence have been considered in the context of general relativity for describing the expansion of the Universe in the background of the de Sitter cosmological metric. The energy-momentum tensor of the most simple zero-mass longitudinal vector fields enters the Einstein equations as a kind of
cosmological constant: therefore the longitudinal waves could really contribute to the dark energy arising \cite{Meierovich}. Another known approach to describe longitudinal spin-1 waves is the St\"uckelberg theory \cite{Stuckelberg}, which recovers the longitudinal degree of freedom lost in the Proca theory for a massive vector field (due to the trasversal Lorentz gauge), in such a way restoring the U$(1)$ gauge invariance.  
Furthermore, in \cite{Perkins} it is shown that a massive vector particle can exist only in the helicity-0 state if it is composed of a fermion-antifermion pair (the mass being generated by the actractive interaction between the two massless particles). In that paper it is also proved that a massive helicity-0 vector particle should behave as a pseudoscalar and that could be identified experimentally by detecting an asymmetry in its decay products, since the longitudinal polarization can lead to forward-backward asymmetries.\footnote{Furthermore, unlike a charged particle with spin 1/2 or spin 1, a charged helicity-0 particle will have zero magnetic moment and its direction of polarization will not be altered by a magnetic field.}

Let us also notice that the Lorentz symmetry of vacuum quantum fields, usually satisfied by Klein--Gordon spin-zero fields (as e.g. the ordinary Higgs boson), \textit{is not broken by HZPs}, because for such particles the quantum vacuum expectation value of the spin vector (equal, as is known, to the helicity) vanishes in any frame. Therefore the helicity-0 field could eventually play the r$\hat{\rm o}$le of a proper quantum vacuum field.

\

\nn In the next section we shall analyze the arising of a nonvanishing spin vector due to a boost applied to the center-of-mass frame (CMF), a sort of ``extrinsic'' spin analogous to the orbital angular momentum.
Then we shall see that, differently from scalar spinless particles, at a classical level HZPs own internal degrees of freedom so that they can be considered as extended-like particles.  Actually, we shall prove that they are special solutions of a very general classical model for free spinning particles corresponding to degenerate oscillatory motions in the CMF.

\section{Extrinsic spin vector}

\nn For a spinning particle with 4-momentum $p^\mu = (p^0; \imp)$, total angular momentun $J^\munu$, spin tensor $S^\munu$, and spin 3-vector $\sbf$ the (conserved) Pauli--Lubanski 4-vector is defined as
\bb
W^\mu \equiv \um\,\vare^{\mu\nu\rho\sigma}J_{\nu\rho}p_\sigma
= \um\,\vare^{\mu\nu\rho\sigma}S_{\nu\rho}p_\sigma
= (\sbf\cdot\imp\,; \; p^0\sbf-\imp\times\kbf)               \label{eq:Wmu}
\ee
where 
$\kbf$ 
is the Lorentz-boost generator (in the particle ``inner'' spin space)
\bb
S_\munu \equiv \left(
\begin{array}{cccc}
0   & \ k^1 & \ k^2 & \ k^3\\
-k^1   & \ 0 & \ -s^3 & \ s^2\\
-k^2   & \ s^3 & \ 0 & \ -s^1\\
-k^3  & \ -s^2 & \ s^1 & \ 0\\
\end{array}
\right)
\ee
The square of $W^\mu$ is a Casimir-invariant of the Poincar\'e group and is equal\footnote{We adopt the signature $+ - - -$ for the metric; hereafter $c=\hbar=1$.} to $-m^2\sbf\cmf^2$, indicating by $\sbf\cmf$ the spin vector in the CMF where $p^\mu = (m;0,0,0)$ \ and \ $W^\mu=(0; m\sbf\cmf)$. \
If in the CMF the spin vector vanishes, $\sbf\cmf=0$, there it vanishes also the Pauli--Lubanski 4-vector, $W^\mu=(0;0,0,0)$. In that case $W^\mu$ \textit{is zero in any reference frame}: 
\bb
\sbf\cdot\imp \, = p^0\sbf-\imp\times\kbf = 0
\label{W=0}
\ee
It follows that in a generic reference frame we shall have
\bb
\sbf = \wbf\times\kbf
\label{spinout}
\ee
where $\wbf\equiv\imp/p^0$ is the center-of-mass speed\footnote{Of course, Eq,\,(\ref{spinout})
can also be obtained from the Lorentz-transformated $S^\munu$, applying a boost $-\wbf$ to the
CMF (as a consequence the particle's center-of-mass, at rest in the CMF, acquires a velocity $\wbf$ in the new frame) and taking into account that $\sbf\cmf=0$
$$
\kbf \ug \gamma\,\kbf\cmf
$$
$$
\sbf \ug \gamma\,\wbf\times\kbf\cmf = \wbf\times\kbf
$$
where as usual $\gamma=1/\sqrt{1-\wbf^2}$ indicates the Lorentz factor.}. Actually the spin $\wbf\times\kbf$ results always orthogonal to $\kbf$\footnote{From the orthogonality between $\sbf$ and $\kbf$ it also follows 
$$
e^{\mu\nu\rho\sigma}S_{\munu}S_{\rho\sigma} \equiv \widetilde{S}S = \sbf\cdot\kbf = 0
$$
For the other Lorentz invariant quantity built up with the spin tensor we find
$$
S^\munu S_\munu = \sbf^2 - \kbf^2 = -\kbf\cmf^2
$$} and to $\wbf$, and then also to the momentum, as expected from the vanishing of the helicity.
We notice that even for $\sbf\cmf=0$, if $\kbf\cmf\neq 0$, the spin vector is not necessarily vanishing in a reference frame other than the CMF: namely, in principle \textit{we can have particles with zero intrinsic angular momentum, i.e. spin zero in the CMF, but with nonzero spin in other reference frames}. This case has not been sufficiently studied in the literature. Notice also that for such particles the momentum is orthogonal to the spin and the helicity is zero, $\sbf\cdot\imp=0$, \textit{in any frame}. Let us remark that, only in this case, that is when also the spatial part of the Pauli-Lubanski 4-vector vanishes, \textit{the helicity is a relativistic invariant quantity}. Hence, in this special case the (zero) \textit{helicity results to be an intrinsic property}, just as spin modulus or electrical charge which classify a given particle and are observer-independent. 

Elementary particles with vanishing intrinsic angular momentum and helicity, then obeying Eq.\,(\ref{W=0}), belong to two different classes:

\begin{itemize}

\item[a)] particles with spin tensor $S^\munu$ identically zero (and then $\sbf=\kbf=0$) in all frames. This is the case of ordinary \textit{scalar spinless} particles described by the Klein--Gordon equation

\item[b)] particles with nonvanishing spin tensor ($S^\munu,\,\sbf,\,\kbf\neq 0$) in a generic reference frame (in the CMF only the polar part of $S^\munu$, i.e. $\kbf$, is different from zero). These helicity-0 particles are non-scalars 
also if, just like Klein--Gordon scalars, are endowed with only one degree of freedom. 

\end{itemize}

\nn In a sense we could say that the particles described in b) seem to own, besides the usual orbital angular momentum $\lbf\equiv\xbf\times\imp$, also a kind of ``extrinsic'' spin vector, which arises only in the presence of an ``external'' motion of the center-of-mass, and disappears in the CMF, just like the orbital momentum does.

\section{Classical spin}

\nn Quantum mechanics of the particles which constitute the ordinary matter (spin--$\um$ particles as quark or electrons) presents a phenomenon known as {\em Zitterbewegung}  (ZBW), described for the first time by Schr\"odinger in the 1930s \cite{Schroedinger,Dirac,Salesi,Classical,Barut,Corben,Papapetrou}: namely, the independence between velocity and momentum (even without external forces)
\bb
v \neq \frac{p}{m}
\ee
Actually in Dirac theory the velocity and momentum operators are not proportional
\bb
\vbfo=\albf\neq\impo=-i\vecna
\ee
Furthermore $\vbfo$, differently from $\impo$, does not commute with the Dirac Hamiltonian $\Hco=\albf\cdot\impo+m\beta$ so that, while $\imp$ is a constant quantity, $\vbf$ is not.
Therefore {\em the dynamical nature of quantum mechanics is intrinsically non-Newtonian}, as it appears more evident after time-derivating side by side the previous equation
\bb
\vbf\neq\frac{\imp}{m} \Impl \abf\neq\frac{\rd\imp}{\rdt}=\frac{\Fbf}{m}
\ee
Thus, for microsystems in general {\em the Newton's Law and (in the absence of external forces) the Galileo's Principle of Inertia do not apply anymore}, and the free motion is not in general uniform rectilinear. The ZBW is fully depicted through the celebrated Gordon deccomposition \cite{Gordon} of the Dirac probability current (here $\hbar=1$):
\bb
j^\mu=\psib\ga^\mu\psi=\frac{1}{2m}\,\lqq\psib(\po^\mu\psi) - (\po^\mu\psib)\psi\rqq +
\frac{1}{m}\pa_\nu\,(\psib\So^\munu\psi)
\ee
($\psib\equiv\psid\gao$, $\po^\mu \equiv i\pa^\mu$, e $\So^{\mu\nu}\equiv
i(\ga^\mu\ga^\nu - \ga^\nu\ga^\mu)/4$ represents the spin tensor operator).
The first term in the r.h.s.\,\,is associated with the ``external'', translational motion of
the center-of-mass; 
whilst, the non-Newtonian term in the r.h.s.\,\,is related to 
the spin, and describes the ``internal'', ZBW rotational motion. Actually, with the separation
between the global motion of the charge and the mean motion of the center-of-mass, a spinning particle appears to be as ``extended-like'', occupying in the CMF a region with dimensions of the order of the Compton radius $1/2m$. This extended-like structure of non-Newtonian particles has an impressive counterpart in the famous ``Darwin term'' which appears in the non-relativistic approximation of the Dirac equation in the presence of an external electromagnetic field, found by C.G. Darwin \cite{Darwin} already in 1928.
Indeed, because of the ZBW the electron undergoes extremely rapid fluctuations on scales just of the order of the Compton wavelength, causing, for example, the electrons moving inside an atom to experience a smeared nuclear Coulomb potential. 

Analogous Gordon-like decompositions of the conserved 4-currents can be found also for spin-1 bosons and for spin-${3\over 2}$ fermions in the Proca and Rarita-Schwinger theories, respectively.

For what seen so far, we presume that any classical (both relativistic and not) theory of spinning particles must predict the ZBW motion, as a signature of the presence of spin freedom 
degrees. In a paper of ours\cite{NNM} it was proposed a {\em classical} particle theory in which spin and ZBW arise quite naturally. For the above considerations we called that theory {\em Non-Newtonian Mechanics} (NNM).
The classical motion of spinning particles was therein described without recourse to particular models or special formalisms, and without employing 
Clifford algebras, or classical spinors (appearing in all supersymmetric-like classical models, as the Barut-Zanghi one\cite{Barut}), but simply by generalizing the usual spinless theory. It was only assumed the invariance with respect to the Poincar\'e group, and only requiring the conservation of the linear and angular momenta we derived the ZBW and the other kinematical properties and motion constraints. Standard Newtonian mechanics\footnote{Hereafter we shall call ``Newtonian'' the ordinary (relativistic or not) theory, where momentum and velocity are parallel vectors.} is re-obtained as a particular case of that theory: namely for spinless systems with no ZBW.

As abovesaid, in the absence of external forces, the Lagrangian shall be invariant with respect to the Poincaré group: 
thus it must be a function of only scalar quantities which do not depend on particle 
position but instead on 4-velocity and, in case, also on its derivatives.
We can built up, also in the presence of external fields, a covariant Lagrangian formulation of the NNM through a straightforward generalization of the Newtonian Lagrangian $\Lczero=\um mv^2-U$ ($v^2\equiv v_\mu v^\mu$) to 
Lagrangians containing 
proper\footnote{Quantity $\tau$ is the proper time, i.e. the CMF time.} 
time-derivatives of the velocity up to the $n$-th order:
\bb 
\Lcn(\tau; x,v,\dov,\ddov,\ldots) \equiv \um mv^2 + \um k_1\dov^2 + \um k_2\ddov^2 + \cdots - U \equiv \sum_{i=0}^n\,\um k_i{v^{(\rm i)}}^2 - U
\label{eq:Lcn}
\ee
where $U(x)$ is a ($\tau$-independent) scalar potential due to external forces, the $k_i$ are
constant scalar coefficients endowed with alternate signs, $k_0=m$, and \ $\vi\equiv\rd^{\rm i}v/\dtau^{\rm i}$\footnote{Alternate signs are requested for the coefficients of the terms appearing in the Lagrangian if we want only stationary solutions and finite oscillatory motions.}.
The coefficients $k_i$ ---which may be chosen equal to zero for $i$ larger than
a given integer, see below--- might be functions of the self-interaction of
the particle and of its mass and charge (cf.\,the infinite-terms equation of the 
self-radiating classical electron or the ``chronon'' theory of the electron\cite{Spinchron},
$k_i$ can be related to the underlying string structure (or membrane or $n$-brane 
structure) of a spinning particle. 
The consequent Euler-Lagrange equation of motion is a constant-coefficients
$n$-th order differential equation, 
\bb
\frac{\partial\Lc}{\partial x} = \frac{\dot{\partial\Lc}}{\partial\dox} -
{\frac{\ddot{\partial\Lc}}{\partial\ddox}} + {\stackrel{\ldots}{\frac{\partial\Lc}{\partial\dddox}}} -
\cdots
\ee
which appears as a generalization of the
Newton's Law $F=ma$, in which non-Newtonian ZBW terms appear:
\bb                                                   
-\,\frac{\pa U}{\pa x_\mu} \ug m\,a^\mu - k_1\,\ddoa^\mu +
k_2\,\ddddoa^\mu - \cdots\equiv
\sum_{i=0}^n\,(-1)^{{\rm i}}k_i\,\aii^\mu\,     \label{eq:GNEq}
\ee
The canonical momentum
$\dis\frac{\pa\Lc}{\pa\dox_\mu} - \frac{\dot{\pa\Lc}}{\pa\ddox_\mu} +
\ddot{\frac{\pa\Lc}{\pa\dddox_\mu}} - \cdots$ \ conjugate to $x^\mu$
writes
\bb
p^\mu = m\,v^\mu - k_1\,\ddov^\mu + k_2\,\ddddov^\mu - \cdots
\equiv\sum_{i=0}^n\,(-1)^{{\rm i}}\,k_i\,{v^{(\rm 2i)}}^\mu\,, \label{eq:ZeroMom}
\ee
from which we get \textit{the ZBW equation} for $\Lcn$:
\bb
v^\mu \ug \frac{p^\mu}{m} + \frac{k_1}{m}\,\ddov^\mu -
\frac{k_2}{m}\,\ddddov^\mu - \cdots = \frac{p^\mu}{m} -
\sum_{i=1}^n\,(-1)^{{\rm i}}\,\frac{k_i}{m}\,{v^{(\rm 2i)}}^\mu\,. \label{eq:LcnZBWeq}
\ee
Let us remark that the above ZBW equation has been derived in \cite{NNM} in a more general form [of which Eq.\,(\ref{eq:LcnZBWeq}) is a particular case] starting from conservation of the linear and angular momenta:
\bb
v^\mu \ug \frac{p^\mu}{m} - \frac{\doS^\munu p_\nu}{m^2}
\ee
In the previously mentioned paper, we also showed that, with the assumption $k_1=-1/4m$, the first-order NNM, namely \nnmuno, is the very classical analogue of the Dirac theory. As a matter fo fact, the non-newtonian classical equations of motion turn out to be identical to the ones holding for the corresponding quantum operators: we can say that the canonical representation of the \nnmuno represents a suitable classical counterpart of the Dirac algebra. The Dirac ZBW motion which yields the expected value $s=\frac{1}{2}$ is lightlike (of course in \nnmuno they are possible also other motions, also not uniform). Actually, the ``inner'' (i.e. in the CMF) motion of the pointlike charge in a plane normal to the spin vector is uniform circular with constant speed $c$, even if the ZBW cycle average speed $w$ (the ``external''center-of-mass speed) results to be, as expected, always subluminal\footnote{Note that for Dirac ZBW uniform motions we have, just as it occurs for the electromagnetic tensor of freely propagating waves in vacuum, \ $S_\munu S^\munu = \widetilde{S}^\munu S_\munu = 0$.}.
Henceforth let us consider free particles, with absence of external forces $U=0$. 
Through the N\"other Theorem, by satisfying the symmetry under 4-rotations,
the classical spin can be univocally defined employing only classical kinematical
quantities.
Actually, requiring Lorentz symmetry for any $\Lcn(\tau; x,v,\dov,\ddov,\ldots)$,
we obtain \cite{NNM} the conservation of the total angular momentum $J^\munu$.
%
Hence, we shall have for the ordinary Newtonian case (the only spinless) $\Lczero$
\bb
J^\munu = \icsmu p^\nu - x^\nu p^\mu
\ee
\bb
\sbf \ug 0
\ee
The spin arises from the first order forward: for $\Lcuno$ we get
\bb
J^\munu = \icsmu p^\nu - x^\nu p^\mu +
k_1\left(v^\mu a^\nu - v^\nu a^\mu\right)       \label{eq:J1}
\ee
hence
\bb
S^\munu\ug k_1\,(v^\mu a^\nu - v^\nu a^\mu) \qquad\qquad \sbf=k_1\,(\vbf\times\abf)               \label{eq:spin1}
\ee
The spin vectors for $\Lcdue$ and $\Lctre$ are, respectively
\bb
\sbf=k_1\,(\vbf\times\abf) + k_2\,(\abf\times\doabf - \vbf\times\ddoabf)
\label{eq:spin2}
\ee
\bb
\sbf=k_1(\vbf\times\abf) + k_2(\abf\times\doabf - \vbf\times\ddoabf) +
 k_3(\doabf\times\ddoabf - \abf\times\dddoabf + \vbf\times\ddddoabf)
\label{eq:spin3}
\ee
For a generic $n$ $(\geq 1)$ spin tensor and spin vector write, respectively:
\bb   
S_\munu \ug \sum_{i=1}^nk_i\,\sum_{l=0}^{i-1}(-1)^{i-l-1}
\left(v_\mu^{(l)}v_\nu^{(2i-l-1)}-v_\nu^{(l)}v_\mu^{(2i-l-1)}\right)
\ee
\bb   
\sbf \ug \sum_{i=1}^nk_i\,\sum_{l=0}^{i-1}(-1)^{i-l-1}
\vbf^{(l)}\times\vbf^{(2i-l-1)}
\label{spinvectorgen}
\ee


\section{Zitterbewegung oscillations with vanishing intrinsic angular momentum and helicity \label{Spinzero}} 

\nn A very interesting case of NNM$^{(n)}$ is the one that, for any $n$, entails only
{\em harmonic motions endowed with only one frequency} $\omz$ with multiplicity $n$:
\bb
\omega_i \ug\omz \qquad\qquad n=1\ldots n
\ee
As is shown in \cite{NNM}, the periodic solution of the ZBW equation of motion (\ref{eq:LcnZBWeq}) endowed with a given frequency $\omega$ can be put as 
\bb
v^\mu(\tau) = \frac{p^\mu}{M} + a^\mu\cos(\omega\tau) + b^\mu\sin(\omega\tau)
\label{vmusol-n}
\ee
Inserting this equation in the ZBW equation 
\bb
v^\mu \ug \frac{p^\mu}{M} - \sum_{i=1}^{n}(-1)^ik_iv_\mu^{(2i)}
\ee
which, reminding that $k_0=M$, can be re-written as
\bb
\sum_{i=0}^{n}(-1)^ik_iv_\mu^{(2i)} \ug \frac{p^\mu}{M}\,,  
\label{harmonic2}
\ee
we get the ``characteristic'' equation in the unknown $\omega$ 
\bb
\sum_{i=0}^{n}k_i\omega^{2i} \ug 0  \label{charact}
\ee
In order that for any solution of the previous equation be  
$$
\omega_i^2 = \omz^2 \qquad\quad i=1,\ldots n
$$
the l.h.s.\,of (\ref{charact}) must be proportional to the polynomial
$F_n$ of grade $n$ in the variable $\omega^2$ defined as follows
\bb
F_n \equiv \sum_{i=0}^{n}\lambda_i\omega^{2i} \equiv
(\omega^2-\omz^2)^n
\ee
where $\lambda_i$ are 
expressed through the Newton binomial coefficients
\bb
\lambda_i \ug \frac{(-1)^{n-i}n!}{i!(n-i)!}\omz^{2(n-i)}
\ee
For the condition $\dis\sum_{i=0}^{n}k_i\omega^{2i} \propto F_n$ \ 
we have to require
\bb
k_i \propto \lambda_i       \label{propto}
\ee
Since $k_0\equiv M$ and $\lambda_0=(-1)^n\omz^{2n}$, from (\ref{propto}) we infer
\bb
k_i \ug \frac{(-1)^nM}{\omz^{2n}}\lambda_i = \frac{(-1)^in!}{i!(n-i)!}\frac{M}{(\omz)^{2i}}
\label{k-lambda}
\ee
               
\nn On inserting solution (\ref{vmusol-n}) with $\omega=\omz$ and coefficients $k_i$ given by Eq.\,(\ref{k-lambda}) in ($n\geq 1$) spin vector formula (\ref{spinvectorgen})
\bb
\sbf = \sum_{i=1}^nk_i\,\sum_{l=0}^{i-1}(-1)^{i-l-1} \vbf^{(l)}\times\vbf^{(2i-l-1)} 
\ee
and referring to CMF ($\imp=0$), 
after some algebra we obtain the spin vector in the following form
\bb
\sbf = \sum^n_{i=1}\frac{(-1)^in!}{(i-1)!(n-i)!}m\rbf\times\vbf
\ee
This expression is different from 0 only for $n=1$ (\nnmuno, i.e.: the Dirac case), whilst for $n\geq 2$ we identically get 
\bb   
\sbf \ug 0
\ee
Therefore, when the intrinsic ---i.e. in the CMF---
motion is harmonic\footnote{For example when it is uniform circular and lightlike with $v^2=0=\const$: cf.\,the classification of ZBW motions proposed in \cite{NNM}.} and is endowed with a degenerate frequency, the spin vector does vanish (only) in CMF, so that \textit{helicity and intrinsic angular momentum are zero}. 
\footnote{Note that even if the spin is zero in the CMF, 
the intrinsic magnetic momentum \ $\dis\um\,e\,\rbf\times\vbf$ \ does not indentically vanish.}

\nn Let us notice a quite interesting analogy between the degenerate frequency of the above seen ZBW inner motion and the degenerate oscillations ($\omega_1=\omega_2$) found in pure fourth-order case of the Pais-Uhlenbeck double oscillator \cite{PU}. As a matter of fact, in \cite{Beltran} it is proved that the degenerate Pais-Uhlenbeck scalar field can be identified with the helicity-0 longitudinal component of the St\"uckelberg field.

\

\nn In addition to the special case of motion with a $n-$degenerate frequency studied above, 
in \cite{NNM} we 
predicted particles endowed with non-zero spin vector and ZBW 
in a generic frame, 
but with zero intrinsic angular momentum and helicity, for any $n\geq 1$ NNM$^{(n)}$ \textit{and for any given coefficients} $k_i$.
It is sufficient to consider, for a chosen $\Lcn$, those solutions of the Euler-Lagrange equation which entail a {\em rectilinear oscillatory motion} in the CMF. To make an example, for $\Lcuno$ it is enough to assume, whichever is the chosen value of $k_1$, 
a rectilinear harmonic\footnote{The linear oscillatory motion is harmonic only for $\Lcuno$; it is in general anharmonic for $n\geq 2$.} motion among the solutions of the ZBW equation (\ref{eq:LcnZBWeq})
\bb
v^\mu = \frac{p^\mu}{m} + F^\mu\,\cos(\omega\tau)
\ee
The space trajectory turns out to be a tilted sinusoid-like path belonging to the plane $\alpha$ containing (as usual we label by $\cmf$ the quantities referred to the CMF) $\vbf\cmf$, $\abf\cmf$ and $\imp$, with the nodal axis parallel to $\imp$. Actually the trajectory is quite different from the one of a spinless Klein--Gordon particle (described by NNM$^{(0)}$), 
which is a straight line.
The spin vector is equal to $k_1\,(\vbf\times\abf)$ because of Eq.\,(\ref{eq:spin1}). By labelling with $\perp$ the component orthogonal to the boost and taking into account 
that, since the Lorentz boost $-\wbf$\footnote{As made in Sect.\,1, we apply an arbitrary boost $-\wbf$ to the CMF and consequently the particle's center-of-mass, at rest in the CMF, acquires a velocity $\wbf$ and a momentum $\imp=p^0\wbf=\gamma m\wbf$ in the new frame.} does not affect the components orthogonal to $\imp$, \ $\Fbf\times\imp=\Fbf^\perp\times\imp=\Fbf^\perp\cmf\times\imp$, we obtain
\bb
\sbf\ug\frac{1}{\sqrt{\omega}}\,(\Fbf\cmf^\perp\times\imp)\,
\sin(\omega\tau)\,.
\ee
Actually the spin does not preceed anymore as the spin of a Dirac particle does, but {\em linearly vibrates along a direction orthogonal to the momentum}, 
and the helicity is zero. 
As expected, in the CMF, where $\imp=0$, the spin vector identically vanishes, 
$\sbf\cmf=0$; while in a generic frame it vanishes only after averaging over a ZBW period, $\overline{\sbf}=0$ (analogously for HZPs it vanishes the quantum average spin, according to the Correspondence Principle).  As is 
easy to check, from eqs.\,(\ref{eq:spin2}-\ref{eq:spin3}) and from the analogous formulae for $n>3$, also for $n>1$ in correspondence to CMF rectilinear motions the intrinsic angular momentum $\sbf\cmf$ does vanish at any time, since $\vbf\cmf$ and its time derivatives are collinear vectors ($\vbf\cmf /\!/\abf\cmf/\!/ \doabf\cmf /\!/ \ \ldots)$. But it easily proved that after a boost the Lorentz-transformed spin vector $\sbf$ is in general non-zero. 
In fact, labelling with $\parallel$ ($\perp$) the components parallel (orthogonal) to the boost $-\wbf$, and taking into account that $v^0\cmf\equiv\dis\frac{\rd\tau}{\rd\tau}=1$  
\ and \ $a^0\cmf=0$\footnote{In the CMF we can write \ $v^\mu=(1;\,\vbf\cmf)$, \ $a^\mu=(0;\,\abf\cmf)$.} we can write for the Lorentz-transformed components of the 4-vectors $v^\mu$ and $a^\mu$
$$
v\para = \gamma\,(v\para\cmf + w)
\qquad \qquad v\orto \ug v\orto\cmf\,,
$$
$$
a\para = \gamma\,a\para\cmf
\qquad \qquad a\orto \ug a\orto\cmf\,,
$$
and so on for the higher-order derivatives of the velocity. As a consequence, in the new frame the transformed velocity $\vbf$ and its transformed derivatives {\em are not anymore collinear}, so that $\sbf\neq 0$.
It is also easy to check that $\vbf$, $\abf$, etc. belong to the aforesaid plane $\alpha$. Being  orthogonal to the plane $\alpha$ for Eqs.\,(\ref{eq:spin1})-(\ref{eq:spin3}) and for the analogous formulae for $n>3$, the spin is then normal to the momentum, 
so that, as expected, the helicity always vanishes.

\

\

\nn {\large\bf Acknowledgements}

\vspace*{0.1cm}

\nn Many thanks are due to E.\,Recami for interesting discussions and useful hints.


\begin{thebibliography}{99} 


\bibitem{Meierovich} B.E. Meierovich: Phys. Review {\bf D84} (2011) 064037 

\bibitem{Stuckelberg} E.C.G. St\"uckelberg: Helv. Phys. Acta {\bf 11} (1938) 225; \textit{ibid.} 225, \textit{ibid.} 312

\bibitem{Perkins} W.A. Perkins: http://arXiv.org hep-ph/0409166 (2004)

\bibitem{Schroedinger} E. Schr\"{o}dinger: Sitzunger. Preuss. Akad. Wiss.
Phys.-Math. Kl. {\bf 24} (1930) 418; {\bf 25} (1931) 1

\bibitem{Dirac} P.A.M. Dirac: {\em The principles of Quantum Mechanics} (Claredon;
Oxford, 1958), $4^{\rm th}$ edition, p.262; \ J. Maddox: Nature  325 (1987) 306


\bibitem{Salesi} G. Salesi: Mod. Phys. Lett. {\bf A11} (1996) 1815;
Int. J. Mod. Phys. {\bf A12} (1997) 5103; \ G. Salesi and E. Recami: Phys. Lett.
{\bf A190} (1994) 137; {\bf A195} (1994) E389; Found. Phys. {\bf 28} (1998)
763; \ E. Recami and G. Salesi: Phys. Rev. {\bf A57} (1998) 98;
Adv. Appl. Cliff. Alg. {\bf 6} (1996) 27; in {\em Gravity, Particles
and Space-Time}, ed. by P.Pronin and G. Sardanashvily (World Scient.; Singapore,
(1996), pp.345-368;
\ G. Cavalleri and G. Salesi: ``$\hbar$ Derived from Cosmology and Origin
of Special Relativity and QED'', in {\em Proceedings of ``Physical
Interpretations of Relativity Theory''} (British Society for the Philosophy of
Science; London, 9--12 September, 1994);
\ M. Pav\v{s}i\v{c}, E. Recami, W.A. Rodrigues, G.D. Maccarrone, F.
Raciti and G. Salesi: Phys. Lett. {\bf B318} (1993) 481; \ W.A. Rodrigues, J.
Vaz, E. Recami and G. Salesi: Phys. Lett. {\bf B318} (1993) 623; \ J. Vaz and
W.A. Rodrigues: Phys. Lett. {\bf B319} (1993) 203

\bibitem{Classical}
\ A.O. Barut: Z. Naturforsch. {\bf A33} (1978) 993;
\ G. Cavalleri: Nuovo Cim. {\bf B55} (1980) 392; Phys. Rev. {\bf D23} (1981)
363; {\bf C6} (1983) 239; Lett. Nuovo Cim. {\bf 43} (1985) 285;
\ M. Mathisson: Acta Phys. Pol. {\bf 6} (1937) 163;
\ H. H\"{o}nl: Ergeb. Exacten Naturwiss. {\bf 26} (1952) 29;
\ K. Huang: Am. J. Phys. {\bf 20} (1952) 479;
\ J. Weyssenhof and A. Raabe: Acta Phys. Pol. {\bf 9} (1947) 7;
\ E.P. Wigner: Ann. Phys. {\bf 40} (1939) 149;
\ M.H.L. Pryce: Proc. Royal Soc. (London) {\bf A195} (1948) 6;
\ T.F. Jordan and M. Mukunda: Phys. Rev. {\bf 132} (1963) 1842;
\ G.N. Fleming: Phys. Rev. B{\bf 139} (1965) 903;
\ M. Pauri: in {\em Group Theoretical Methods in Physics}, Lectures Notes in
Physics, vol.135, p.615, ed. by J.Ehlers, K.Hepp, R.Kippenhahn,
H.A.Weidenm\"uller and J.Zittartz (Springer-Verlag; Berlin, 1980)

\bibitem{Barut} A.O. Barut and N. Zanghi: Phys. Rev. Lett. {\bf 52} (1984) 2009;
\ A.O. Barut and A.J. Bracken: Phys. Rev. {\bf D23} (1981) 2454; {\bf D24}
(1981) 3333; \ A.O. Barut and I.H. Duru: Phys. Rev. Lett. {\bf 53} (1984) 2355;
\ A.O. Barut and M. Pav\v{s}i\v{c}: Class. Quantum Grav. {\bf 4} (1987) L131;
Phys. Lett. {\bf B216} (1989) 297

\bibitem{Corben} M.J. Bhabha and H.C. Corben: Proc. Roy. Soc. (London) {\bf
A178} (1941) 273; \ H.C. Corben: Phys. Rev. {\bf 121} (1961) 1833;
{\em Classical and Quantum Theories of Spinning Particles} (Holden-Day; San
Francisco, 1968); Phys. Rev. {\bf D30} (1984) 2683; Am. J. Phys. {\bf 45}
(1977) 658; {\bf 61} (1993) 551; Int. J. Theor. Phys. {\bf 34} (1995) 19

\bibitem{Papapetrou} A. Papapetrou: Proc. Roy. Soc. (London) {\bf A209} (1951)
248; \ H. H\"{o}nl and A. Papapetrou: Z. Phys. {\bf 112} (1939) 512; {\bf 116}
(1940) 153

\bibitem{Gordon} W. Gordon: Z. Phys. {\bf 50} (1928) 630; \ J.D.Bjorken and
S.D.Drell: {\em Relativistic Quantum Mechanics}, (McGraw--Hill; New York,
1964), p.35

\bibitem{Darwin} C.G. Darwin, Proc. R. Soc. Lond. {\bf A118} (1928) 654 

\bibitem{NNM} G. Salesi: Int. J. Mod. Phys. {\bf A17} (2002) 347

\bibitem{Spinchron} E. Recami and G. Salesi: Found.Phys. {\bf 37} (2007) 277 

\bibitem{PU} A. Pais and G.E. Uhlenbeck: Phys. Rev. {\bf 79} (1950) 145 

\bibitem{Beltran} J. Beltrán Jiménez, E. Di Dio and R. Durrer: JHEP \textbf{1304} (2013) 030

%
%
%
%
%
%
%
%
%
%

\end{thebibliography}
\end{document}